\documentclass[a4paper,english,aps,reprint]{revtex4-1}
\usepackage[T1]{fontenc}
\usepackage[utf8]{inputenc}
\usepackage{babel}
\usepackage{amsmath}
\usepackage{amssymb}
\usepackage{stackrel}
\usepackage{graphicx}
\usepackage{esint}
\usepackage[unicode=true,pdfusetitle,
 bookmarks=true,bookmarksnumbered=false,bookmarksopen=true,bookmarksopenlevel=1,
 breaklinks=false,pdfborder={0 0 0},backref=false,colorlinks=false]
 {hyperref}

\makeatletter

\pdfpageheight\paperheight
\pdfpagewidth\paperwidth

\@ifundefined{textcolor}{}
{%
 \definecolor{BLACK}{gray}{0}
 \definecolor{WHITE}{gray}{1}
 \definecolor{RED}{rgb}{1,0,0}
 \definecolor{GREEN}{rgb}{0,1,0}
 \definecolor{BLUE}{rgb}{0,0,1}
 \definecolor{CYAN}{cmyk}{1,0,0,0}
 \definecolor{MAGENTA}{cmyk}{0,1,0,0}
 \definecolor{YELLOW}{cmyk}{0,0,1,0}
}

\usepackage{babel}
\usepackage{babel}

\usepackage{babel}
\usepackage[normalem]{ulem}
\usepackage{babel}
\usepackage{babel}
\usepackage{amsfonts}
\providecommand{\U}[1]{\protect\rule{.1in}{.1in}}

\pdfpageheight\paperheight
\pdfpagewidth\paperwidth

\@ifundefined{textcolor}{}{
\definecolor{BLACK}{gray}{0}
\definecolor{WHITE}{gray}{1}
\definecolor{RED}{rgb}{1,0,0}
\definecolor{GREEN}{rgb}{0,1,0}
\definecolor{BLUE}{rgb}{0,0,1}
\definecolor{CYAN}{cmyk}{1,0,0,0}
\definecolor{MAGENTA}{cmyk}{0,1,0,0}
\definecolor{YELLOW}{cmyk}{0,0,1,0}
}

\makeatother

\begin{document}

\title{Improved one-dimensional model potentials for strong-field simulations}

\author{Szilárd Majorosi\textsuperscript{1}, Mihály G. Benedict\textsuperscript{1}
and Attila Czirják\textsuperscript{1,2,}}

\email{czirjak@physx.u-szeged.hu}

\affiliation{\textsuperscript{1}Department of Theoretical Physics, University
of Szeged, Tisza L. krt. 84-86, H-6720 Szeged\linebreak{}
 \textsuperscript{2}ELI-ALPS, ELI-HU Non-Profit Ltd., Dugonics tér
13, H-6720 Szeged, Hungary}
\begin{abstract}
Based on a plausible requirement for the ground state density, we
introduce a novel one-dimensional (1D) atomic model potential for
the 1D simulation of the quantum dynamics of a single active electron
atom driven by a strong, linearly polarized few-cycle laser pulse.
The form of this density-based 1D model potential also suggests improved
parameters for other well-known 1D model potentials. We test these
1D model potentials in numerical simulations of typical strong-field
physics scenarios and we find an impressively increased accuracy of
the low-frequency features of the most important physical quantities.
The structure and the phase of the high-order harmonic spectra also
have a very good match to those resulting from the three-dimensional
simulations, which enables to fit the corresponding power spectra
with the help of a simple scaling function. 
\end{abstract}
\maketitle

\section{Introduction\label{sec:introduction}}

Atomic and molecular physics has witnessed a revolution due to the
appearance of attosecond pulses \cite{hentschel2001attosecond,kienberger2002attosecond,drescher2002attosecond,baltuska2003attosecond,Uiberacker_Nature_2007,Krausz_RevModPhys_2009_Attosecond_physics,hommelhoff2009extremelocalization,Schultze_Science_2010,Haessler_NatPhys_2010,pfeiffer2012attoclock,shafir2012tunneling,ranitovic2014attosecondcontrol,peng2015attosecondtracing,ciappina2017attosecondnano}.
The true understanding of the phenomena in attosecond and strong-field
physics often needs the quantum evolution of an involved atomic system
driven by a strong laser pulse \cite{Keldysh_JETP_1965,varro1993multiphotonequation,lewenstein1994hhgtheory,protopapas1997tdseionization,ivanov2005strongfield,gordon2005tdsecoulomb,frolov2012attosecondanalytic}.
However, analytical or even numerically exact solution of the corresponding
Schrödinger equation is beyond reach in this non-perturbative range,
except for the simplest cases. Therefore, approximations are unavoidable
and very important.

For linearly polarized pulses, the main dynamics happens along the
electric field of the laser pulse which underlies the success of some
one-dimensional (1D) approximations \cite{eberly1988softcoulombspecra,eberly1991softcoulombatom,bauer1997tdse1dhemodel,chirilua2010HHGemission,silaev2010tdsecoulombs,sveshnikov2012schrodingercoulomb,graefe2012quantumphasespace,czirjak2000ionizationwigner,czirjak2013rescatterentanglement,geltman2011boundstatesdelta,baumann2015wignerionization,teeny2016ionizationtime}.
These typically use various 1D model potentials to account for the
behavior of the atomic system. However, the particular model potential
chosen heavily influences the 1D results and their comparison with
the true three-dimensional (3D) results is usually non-trivial. One
of these important deviations is that the dipole moment, created by
the same electric field, may have much larger or much smaller values
in the 1D than in the 3D simulation.

In the present paper, we introduce and test novel 1D atomic model
potentials for strong-field dynamics driven by a linearly polarized
laser pulse. Our key idea is to require the ground state density of
the 1D model to be equal to the reduced 3D ground state density, obtained
by integrating over spatial coordinates perpendicular to the direction
of the laser polarization. According to density functional theory,
this 1D ground state density determines the corresponding 1D model
potential up to a constant, which we set by matching the ground state
energies. Comparison of the resulting new formula with well-known
1D model potentials inspires us to use some of the latter with improved
parameters. Then we test these improved 1D model potentials by applying
them in careful numerical simulations of strong-field ionization by
a few-cycle laser pulse. Based on these results we make a conclusion
about the best of these novel 1D model potentials. We use atomic units
in this paper.

\section{3D and 1D model systems\label{sec:3D-and-1D-model-systems}}

\subsection{3D reference system\label{sec:3D-reference-system}}

First, we define the 3D system which we aim to model in 1D. We write
the Hamiltonian $H_{0}^{{\rm 3D}}$ of a 3D hydrogen atom or hydrogen-like
ion in cylindrical coordinates $\rho=\sqrt{x^{2}+y^{2}}$ and $z$
as 
\begin{equation}
H_{0}^{{\rm 3D}}=T_{z}+T_{\rho}-\frac{Z}{\sqrt{\rho^{2}+z^{2}}}\label{eq:3d_hamiltonian_0}
\end{equation}
where $Z$ is the charge of the ion-core ($Z=1$ for hydrogen) and
the two relevant terms of the kinetic energy operator are given by
\begin{equation}
T_{\rho}=-\frac{1}{2\mu}\left[\frac{\partial^{2}}{\partial\rho^{2}}+\frac{1}{\rho}\frac{\partial}{\partial\rho}\right],\,\,\,\,\,\,\,\,\,\, T_{z}=-\frac{1}{2\mu}\frac{\partial^{2}}{\partial z^{2}},\label{eq:3d_kinetic_z_rho}
\end{equation}
where $\mu$ denotes the (reduced) mass of the reduced system. By
solving the equation 
\begin{equation}
H_{0}^{{\rm 3D}}\psi_{100}(z,\rho)=E_{0}\psi_{100}(z,\rho)\label{eq:3d_schroedinger}
\end{equation}
we get the well known ground state energy and wavefunction of the
Coulomb problem \cite{BOOK_QUANTUM_GRIFFITS_2005,BOOK_ATOMS_MOLECULES_BRANSDEN_2003}
as 
\begin{equation}
E_{0}=-\frac{\mu Z^{2}}{2},\,\,\,\,\,\,\,\,\,\,\psi_{100}(z,\rho)=\mathcal{N\,}e^{-\mu Z\sqrt{\rho^{2}+z^{2}}},\label{eq:3d_coulomb_eigensys}
\end{equation}
where $\mathcal{N}$ is a real normalization factor. We consider the
action of a linearly polarized laser pulse on this atomic system in
the dipole approximation by the potential 
\begin{equation}
V_{\text{ext}}(z,t)=z\cdot\mathcal{E}_{z}(t),\label{eq:pot1_length_z}
\end{equation}
and seek solutions of the time-dependent Schrödinger equation 
\begin{equation}
i\frac{\partial}{\partial t}\Psi^{{\rm 3D}}\left(z,\rho,t\right)=\left[H_{0}^{{\rm 3D}}+V_{\text{ext}}(z,t)\right]\Psi^{{\rm 3D}}\left(z,\rho,t\right)\label{eq:3d_tdse}
\end{equation}
that start from the $\psi_{100}(z,\rho)$ ground state at $t=0$,
and we compute it up to a specified time $T_{\max}$. This time-dependent
Hamiltonian still has axial symmetry around the direction of the electric
field of the laser pulse which makes the use of cylindrical coordinates
practical. For the efficient numerical solution of the time evolution
in real space, we use the algorithm described in \cite{majorosi2016tdsesolve}
which incorporates the singularity of the Hamiltonian directly, using
the required discretized Neumann and Robin boundary conditions.

\subsection{1D model system\label{subsec:1D-model-system}}

In order to model the above described 3D time-dependent process in
1D, it is customary to assume a 1D atomic Hamiltonian of the following
form: 
\begin{equation}
H_{0}^{{\rm 1D}}=T_{z}+V_{0}^{{\rm 1D}}(z)\label{eq:1d_hamiltonian_0}
\end{equation}
where $V_{0}^{{\rm 1D}}(z)$ is an atomic model potential of choice,
and then to seek solutions of the time-dependent Schrödinger equation
\begin{equation}
i\frac{\partial}{\partial t}\Psi^{{\rm 1D}}\left(z,t\right)=\left[H_{0}^{{\rm 1D}}+V_{\text{ext}}(z,t)\right]\Psi^{{\rm 1D}}\left(z,t\right)\label{eq:1d_tdse}
\end{equation}
where the external potential $V_{\text{ext}}(z,t)$ is given in \eqref{eq:pot1_length_z}\@.
In this article we are going to introduce a new form of $V_{0}^{{\rm 1D}}(z)$
to model strong-field processes physically as correctly as possible.
But before doing so, let us shortly recall some of the 1D potentials
used earlier. We shall then propose certain improvements of these
known formulas aiming that the resulting 1D simulations reproduce
the 3D system's strong-field response quantitatively correctly.

\subsection{Conventional 1D model potentials\label{subsec:Conventional}}

There are a number of well-known 1D atomic model potentials in the
literature \cite{silaev2010tdsecoulombs,sveshnikov2012schrodingercoulomb,graefe2012quantumphasespace},
having their advantages and disadvantages. Here we summarize the basics
of two of these, which we think to be the most important for the modeling
of strong-field phenomena.

The soft-core Coulomb potential is defined as 
\begin{equation}
V_{0,{\rm Sc}}^{{\rm 1D}}(z)=-\frac{Z}{\sqrt{z^{2}+\alpha^{2}}}\label{eq:pot1_sc}
\end{equation}
where the smoothing parameter $\alpha$ is usually adjusted to match
the ground state energy to a selected single electron energy. For
$\mu=1$, $Z=1$, and $\alpha^{2}=2$, its ground state energy and
ground state can be used as a 1D model hydrogen atom:

\begin{equation}
E_{0,{\rm Sc}}=-\frac{1}{2},\,\,\,\,\,\,\psi_{0,{\rm Sc}}(z)=\mathcal{N}_{{\rm Sc}}\left(1+\sqrt{z^{2}+2}\right)e^{-\sqrt{z^{2}+2}}\label{eq:pot1_sc_groundsys}
\end{equation}
where $\mathcal{\mathcal{N}}_{{\rm Sc}}$ is the normalization factor.
The most important features of this model potential are that it is
a smooth function, it has an asymptotic Coulomb form and Rydberg continuum.
The energy of its first excited state is $E_{1,{\rm Sc}}=-0.2329034$.

The 1D Dirac-delta potential \cite{czirjak2000ionizationwigner,geltman2011boundstatesdelta,czirjak2013rescatterentanglement}
\begin{equation}
V_{0,{\rm DD}}^{{\rm 1D}}(z)=-Z\delta(z)\label{eq:pot1_dd}
\end{equation}
has the following ground state energy and ground state: 
\begin{equation}
E_{0,{\rm DD}}=-\frac{Z^{2}}{2},\,\,\,\,\,\,\,\,\,\,\psi_{0,{\rm DD}}(z)=\sqrt{Z}e^{-Z|z|}\label{eq:pot1_dd_groundsys}
\end{equation}
for $\mu=1$. The singularity of $V_{0,{\rm DD}}^{{\rm 1D}}(z)$ at
$z=0$ is sometimes considered as a disadvantage, but this gives rise
to a Robin boundary condition, just like the Coulomb singularity does
in 3D. Hence, this potential has its ground state with the same exponential
form and cusp, and ground state energy as that of the 3D hydrogen
atom (with $Z=1$).

Despite these facts, the experience shows that \eqref{eq:pot1_sc}
and \eqref{eq:pot1_dd} do not give strong field simulation results
that would be quantitatively comparable to those of the reference
3D system (cf. \cite{bandrauk2009tdsehydrogen,graefe2012quantumphasespace,majorosi2017entanglement})
therefore the model system parameters need to be manually adjusted,
for example by changing the strength of $V_{\text{ext}}(z,t)$.

\section{Density based model potentials\label{sec:density based model}}

\subsection{Derivation of the 1D analytical model potential\label{sub:pot1_model_analytical}}

We are going to derive now a new formula for $V_{0}^{{\rm 1D}}(z)$,
and based on its peculiarities we will suggest certain improvements
in other known 1D model potentials.

Our inspiration of deriving this new 1D model potential originated
from the ground state density functional theory of multielectron atoms.
More specifically, the following derivation is analogous to the derivation
of the Kohn-Sham potential of a helium atom with a single orbital:
knowing the correct reduced (single particle) density \cite{BOOK_TDDFT_ULLRICH}
one can invert the Schrödinger equation to determine the Kohn-Sham
potential \cite{kohn1965dftequations}. In this way one can model
the ground state of the system as accurately as it is possible with
a single orbital. However, in the present paper we consider just single
active electron atoms and we make the analogous reduction from the
3D electron coordinates to the $z$ coordinate of the single electron.

\begin{figure}[h]
\includegraphics[width=1\columnwidth]{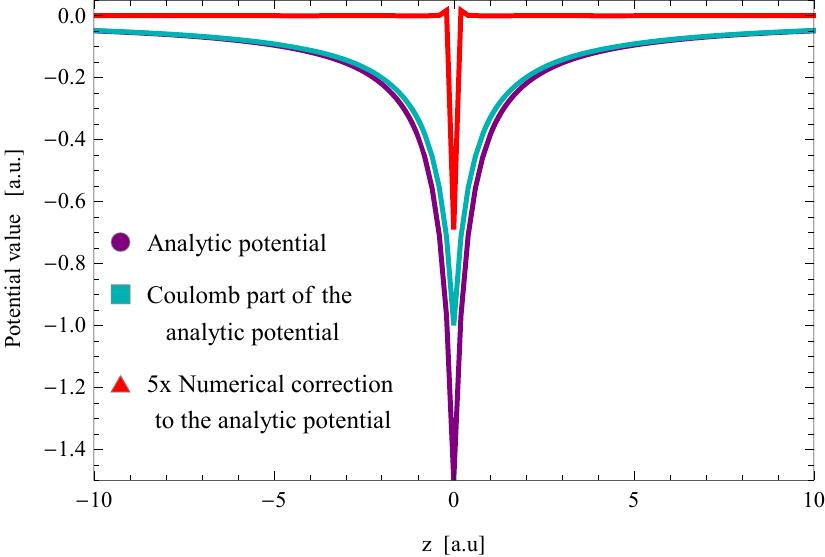}\protect\protect\caption{Plot of the analytic potential \eqref{eq:pot1_model} (in purple)
and its regularized Coulomb part (second term of \eqref{eq:pot1_model},
in cyan), for $Z=1$, $\mu=1$. We also plot the difference $\widetilde{V}_{0,{\rm M}}^{{\rm 1D}}(z)-V_{0,{\rm M}}^{{\rm 1D}}(z)$,
(see the discussion in section \ref{sec:numerical}) calculated with
for $\Delta z=0.2$, and magnified by a factor of 5 (in red). This
is to illustrate the numerical correction to be introduced by Eq.
\eqref{eq:pot1_model_num}. \label{fig:z_potential_plot}}
\end{figure}

For developing our 1D model potential, we need the 1D reduced density
of the 3D ground state that is defined by 
\begin{equation}
\varrho_{z}^{100}(z)=2\pi\int_{0}^{\infty}|\psi_{100}(z,\rho)|^{2}\rho\text{d}\rho.\label{eq:z_density}
\end{equation}
After the substitution of \eqref{eq:3d_coulomb_eigensys} for the
integrand, we can perform this integral analytically which yields
the closed form 
\begin{equation}
\varrho_{z}^{100}(z)=\tfrac{\mu Z}{2}\left(2Z\mu|z|+1\right)e^{-2Z\mu|z|}.\label{eq:z_density_coulomb}
\end{equation}
Our key idea is now to require the 1D model system to have its ground
state density identical with $\varrho_{z}^{100}(z)$. According to
density functional theory, this 1D ground state density determines
the corresponding 1D model potential $V_{0,{\rm M}}^{{\rm 1D}}(z)$
up to a constant. We can calculate this potential straightforwardly:
we define the ground state of the 1D model atom obviously as $\psi_{0}(z)=\sqrt{\varrho_{z}^{100}(z)}$,
i.e. 
\begin{equation}
\psi_{0}(z)=\sqrt{\tfrac{\mu Z}{2}}\sqrt{2\mu Z|z|+1}e^{-\mu Z|z|}\label{eq:pot1_model_state}
\end{equation}
and then we invert the eigenvalue equation of $H_{0}^{{\rm 1D}}$
as 
\begin{equation}
V_{0,{\rm M}}^{{\rm 1D}}(z)=E_{0,{\rm M}}+\frac{1}{\psi_{0}(z)}\frac{1}{2\mu}\frac{\partial^{2}}{\partial z^{2}}\psi_{0}(z).\label{eq:pot1_model_invert}
\end{equation}
After performing the differentiation we get 
\begin{equation}
V_{0,{\rm M}}^{{\rm 1D}}(z)=E_{0,{\rm M}}+\frac{2\mu^{3}Z^{4}|z|^{2}-\mu Z^{2}}{\left(2\mu Z|z|+1\right)^{2}}.\label{eq:pot1_model_an}
\end{equation}
In order to determine the ground state energy, we rewrite this potential
as 
\begin{equation}
V_{0,{\rm M}}^{{\rm 1D}}(z)=E_{0,{\rm M}}+\frac{\mu Z^{2}}{2}\frac{\left(2\mu Z|z|+1\right)\left(2\mu Z|z|-1\right)-1}{\left(2\mu Z|z|+1\right)^{2}},
\end{equation}
and then we impose the asymptotic value 
\begin{equation}
\lim_{|z|\rightarrow\infty}V_{0,{\rm M}}^{{\rm 1D}}(z)=0,\label{eq:pot1_asymptotic}
\end{equation}
which yields the ground state energy 
\begin{equation}
E_{{\rm 0,M}}=E_{0}=-\frac{\mu Z^{2}}{2}.\label{eq:pot1_model_energy}
\end{equation}
Using this value, after some algebraic manipulations we arrive at
the following instructive form of our new density-based 1D atomic
model potential:

\begin{equation}
V_{0,{\rm M}}^{{\rm 1D}}(z)=-\frac{1}{2\mu}\frac{1}{2^{2}\left(|z|+\frac{1}{2\mu Z}\right)^{2}}-\frac{\frac{1}{2}Z}{|z|+\frac{1}{2\mu Z}}.\label{eq:pot1_model}
\end{equation}
Let us make a few important notes. It is the asymptotic tail of the
reduced 1D ground state density $\varrho_{z}^{100}(z)$ that determines
the ground state energy $E_{{\rm 0,M}}$ in such a non-trivial way
that it is identical to the ground state energy of the 3D system,
$E_{{\rm 0}}$. The asymptotic tail of $\varrho_{z}^{100}(z)$ also
determines the regularized 1D Coulomb potential with effective charge
$\frac{1}{2}Z$ which is the second term in \eqref{eq:pot1_model}.
This term is dominant over the short range first term of \eqref{eq:pot1_model}
not only in the asymptotic tail but also around the center at least
by a factor of 2, see the corresponding curves of Fig. \ref{fig:z_potential_plot}.
The minima of both of the two terms of $V_{0,{\rm M}}^{{\rm 1D}}(z)$
at $z=0$ decrease with increasing $Z$ or $\mu$. For $Z=1$ and
$\mu=1$, the energy of its first excited state is $E_{1,{\rm M}}=-0.0904408$
approximately. 

\subsection{Improved 1D model potentials}

The results of Sec. \ref{sub:pot1_model_analytical}, especially the
somewhat surprising value of an effective charge of $\frac{1}{2}Z$,
suggested by the second term of the analytical model potential \eqref{eq:pot1_model},
inspire us to use the 1D soft-core Coulomb potential and a 1D regularized
Coulomb potential with accordingly modified values of their parameters.
As we will see, these modifications lead indeed to improved results
in strong-field simulations.

We use $\frac{1}{2}Z$ in the nominator of the soft-core Coulomb potential,
which then requires to change also the parameter $\alpha$ in order
to maintain that its ground state energy matches the 3D ground state
energy. These lead us to the following formula of the improved 1D
soft-core Coulomb potential with $\mu=1$: 
\begin{equation}
V_{0,{\rm M,Sc}}^{{\rm 1D}}(z)=-\frac{\frac{1}{2}Z}{\sqrt{z^{2}+\frac{1}{4Z^{2}}}}\text{ with }E_{{\rm 0,M,Sc}}=-\frac{Z^{2}}{2},\label{eq:pot1_sc_corr}
\end{equation}
that has the correct $\frac{1}{2}Z/|z|$ asymptotic behavior when
$|z|\rightarrow\infty$. The energy of its first excited state is
$E_{1,{\rm M,Sc}}=-0.1058670$.

We also introduce the improved 1D regularized Coulomb potential as
\begin{equation}
V_{0,{\rm M,C}}^{{\rm 1D}}(z)=-\frac{\frac{1}{2}Z}{|z|+a},\label{eq:pot1_cc_corr}
\end{equation}
where the value of the parameter $a$ is determined by requiring that
the ground state energy is $E_{{\rm 0,M,C}}=-\frac{\mu Z^{2}}{2}$.
For $Z=1$ we set $a\approx0.32889$ which yields $E_{{\rm 0,M,C}}\approx-0.5000007$
(for $\mu=1$). We note that this has been computed numerically with
the spatial step size $\Delta z=0.2$. 

The sophisticated numerical method to be outlined in the next section
will enable us to make additional refinements regarding the 1D density
based potential, as well as the 1D delta potential. These improvements
will be explained below especially by the formulas given in Eqs. \eqref{eq:pot1_model_num}
and \eqref{eq:pot1_dd_num}.

\section{Numerical methods of the solution\label{sec:numerical}}

Usually, the time-dependent Schrödinger equation \eqref{eq:1d_tdse}
must be solved numerically in the non-perturbative regime. We discretize
the time variable with time steps $\Delta t$ as $t_{k}=k\Delta t$,
and the spatial coordinate with steps $\Delta z$ as $z_{j}=j\Delta z$
($k,j$ are integer indices). The discretized wave function is written
as $\Psi^{{\rm 1D}}(z_{j},t_{k})$. We write the discretized form
of the 1D atomic model Hamiltonian as 
\begin{equation}
\widetilde{H}_{0}=\widetilde{T}_{z}+V_{0}^{{\rm 1D}}(z_{j}),\label{eq:1d_discrete_h0}
\end{equation}
where, based on our experiences detailed in \cite{majorosi2016tdsesolve,majorosi2017entanglement},
we use the following 11-point finite difference method \cite{dijk2007tdseaccurate}
for the discretization of the kinetic energy operator $T_{z}$: 
\begin{equation}
\widetilde{T}_{z}\Psi(z_{j},t_{k})=-\frac{1}{2\mu}\sum_{s=-5}^{5}c_{|s|}^{(5)}\Psi^{{\rm 1D}}(z_{j}+s\Delta z,t_{k}),\label{eq:1d_discrete_Tz}
\end{equation}
see Table 1 of \cite{dijk2007tdseaccurate} for the coefficients $c$.
This is accurate up to $\Delta z^{10}$ for smooth functions (it is
also limited by the Fourier representation). Then, the discrete Hamiltonian
becomes an 11-banded diagonal matrix which operates on the column
vector of the discretized wave function in coordinate representation.
Regarding the use of the atomic model potential in numerical simulations,
this is the most important step since it defines the numerical eigensystem
of the atom.

Regarding the time evolution, we use a 3 step splitting of the $U(t,t+\Delta t)$
evolution operator which has an accuracy of $\Delta t^{4}$, and each
of its substeps are propagated using the usual second order Crank-Nicolson
method \cite{BOOK_NUMERICAL_RECIPIES_2007} with a discrete second
order effective Hamiltonian, the particular formulas can be found
in Sec 3.1 and Sec 4.1 in \cite{majorosi2016tdsesolve}. We find ground
states and ground state energies performing imaginary time propagation
\cite{bauer2006tdseqprop,chin2009imagsplit}. For integrations, we
use the quadrature formula $\int f(z){\rm d}z\approx\sum_{j}f(z_{j})\Delta z$
because the numerical time evolution is unitary with respect to this
summation.

When using the potential \eqref{eq:pot1_sc_corr} the method described
above can be applied without further complications. In the case of
our density-based model potential a refinement is necessary as \eqref{eq:pot1_model}
is not differentiable in the origin, just as the true 3D Coulomb potential.
Therefore the ground state and energy of the discrete Hamiltonian
\eqref{eq:1d_discrete_h0} with $V_{0,{\rm M}}^{{\rm 1D}}(z_{j})$
is accurate only up to $\Delta z^{2}$. This is the reason why its
ground state density does not equal $\varrho_{z}^{100}(z_{j})$ accurately
enough, unless $\Delta z$ is extremely small. We avoid this inaccuracy
in the following way: instead of $V_{0,{\rm M}}^{{\rm 1D}}(z_{j})$,
we use its following discretized form in the computations:

\begin{equation}
\widetilde{V}_{0,{\rm M}}^{{\rm 1D}}(z_{j})=E_{0}-\frac{1}{\psi_{0}(z_{j})}\widetilde{T}_{z}\psi_{0}(z_{j}).\label{eq:pot1_model_num}
\end{equation}
This definition of $\widetilde{V}_{0,{\rm M}}^{{\rm 1D}}(z_{j})$
ensures that the discretized ground state vector $\psi_{0}(z_{j})$
is the eigenvector of \eqref{eq:1d_discrete_h0} with $\widetilde{V}_{0,{\rm M}}^{{\rm 1D}}(z_{j})$
and the corresponding energy is $\widetilde{E}_{0,{\rm M}}=E_{0}$,
numerically exactly. The energy of the corresponding first excited
state (with $\Delta z=0.2$) is $\widetilde{E}_{1,{\rm M}}=-0.0904385$,
which is close enough to $E_{1,{\rm M}}$. We have plotted the difference
$\widetilde{V}_{0,{\rm M}}^{{\rm 1D}}(z)-V_{0,{\rm M}}^{{\rm 1D}}(z)$
in Fig. \ref{fig:z_potential_plot}, magnified by a factor of 5.

The discretized form of the analytical model potential, $\widetilde{V}_{0,{\rm M}}^{{\rm 1D}}(z_{j})$
suggests also a modified discretization of the Dirac-delta potential
that we introduce as 
\begin{equation}
\widetilde{V}_{0,{\rm DD}}^{{\rm 1D}}(z_{j})=E_{0,{\rm DD}}-\frac{1}{\psi_{0,{\rm DD}}(z_{j})}\widetilde{T}_{z}\psi_{0,{\rm DD}}(z_{j})\label{eq:pot1_dd_num}
\end{equation}
using the corresponding exact ground state $\psi_{0,{\rm DD}}(z_{j})$
and energy $E_{0,{\rm DD}}$. This is a finite discretized potential
which eliminates any singular feature from the corresponding Hamiltonian
matrix. As we show it in Appendix \ref{sec:appendix_accuracy}, such
definitions enable consistent and accurate simulations with high order
finite differences, therefore it is a valid choice to define a potential
using numerical inversion from its ground state.

\section{Results and comparison of the 1D and 3D calculations\label{sec:results}}

In this section, we present and compare the results of strong-field
simulations based on the 1D model potentials discussed in the previous
sections. We selected the mean value of the dipole moment $\left\langle z\right\rangle (t)$
and its standard deviation $\sigma_{z}(t)$, the mean value of the
velocity $\left\langle v_{z}\right\rangle (t)$, and the ground state
population loss $g(t)$ to characterize the dynamics resulting from
the solutions of \eqref{eq:1d_tdse} with the various model potentials
and from the solution of \eqref{eq:3d_tdse} as a reference. We also
investigate the relation between the resulting various dipole power
spectra $p(f)$, which is one of the most important quantities for
high order harmonic generation \cite{McPherson_JOSAB_1987_HHG,Harris_OptCom_1993_HHG_Atto,lewenstein1994hhgtheory}
and attosecond pulses. For the formulas of these physical quantities
and for some details about the numerical accuracy of the simulations,
see Appendix \ref{sec:comparable-physical-quantities} and \ref{sec:appendix_accuracy}.

In these simulations, we model the linearly polarized few-cycle laser
pulse with a sine-squared envelope function. The corresponding time-dependent
electric field has non-zero values only in the interval $0\leq t\leq N_{\text{Cycle}}T$
according to the formula: 
\begin{equation}
\mathcal{E}_{z}(t)=F\cdot\sin^{2}\left(\frac{\pi t}{2N_{\text{Cycle}}T}\right)\cos\left(\frac{2\pi t}{T}\right),\label{eq:sim_sinpulse3_E}
\end{equation}
where $T$ is the period of the carrier wave, $F$ is the peak electric
field strength and $N_{\text{Cycle}}$ is the number of cycles under
the envelope function. Unless otherwise stated, we set $N_{\text{Cycle}}=3$
and $T=100$, the latter corresponds to a ca. $725\,\mathrm{nm}$
near-infrared carrier wavelength. From Fig. \ref{fig:SinPulse3_F010_MeanZ}
on, the vertical dashed lines denote the zero crossings of the respective
$\mathcal{E}_{z}(t)$ electric field.

We consider hydrogen in most of the simulations, i.e. we use $Z=1$
and $\mu=1$ if not otherwise stated explicitly. We set typically
$\Delta z=0.2$ and $\Delta t=0.01$ since these are sufficient for
the numerical errors to be within line thickness. We use box boundary
conditions and we set the size of the box to be sufficiently large
so that the reflexions are kept below $10^{-8}$ atomic units.

The 3D reference results (i.e. the simulation results of the true
3D Schrödinger equation \eqref{eq:3d_tdse}) are plotted in Figs.
\ref{fig:SinPulse3_F010_MeanZ} to \ref{fig:SinPulse3_Four_F015_NE}
in blue and are labeled ``3D reference''. The 1D simulation results
and their respective colors are plotted as follows: our density-based
model potential from numerical inversion \eqref{eq:pot1_model_num}
in purple, our improved soft-core Coulomb potential \eqref{eq:pot1_sc_corr}
in gold, our improved regularized Coulomb potential \eqref{eq:pot1_cc_corr}
in red, the conventional soft-core Coulomb potential \eqref{eq:pot1_sc}
in green and the discretized Dirac-delta potential \eqref{eq:pot1_dd}
in dark blue.

\subsection{Low frequency response}

First, we discuss the results of a moderately strong laser pulse having
a peak electric field value of $F=0.1$. We plot the corresponding
time-dependent mean values $\left\langle z\right\rangle (t)$ (the
magnitude of which equals the dipole moment in atomic units) and their
standard deviations $\sigma_{z}(t)$ in Fig. \ref{fig:SinPulse3_F010_MeanZ},
the time-dependent mean velocities $\left\langle v_{z}\right\rangle (t)$
and the ground state population losses $g(t)$ in Fig. \ref{fig:SinPulse3_F010_Proj}
for all the 1D model systems listed above. These curves justify that
the simulation results obtained with our density-based model potential
and the improved model potentials are already quantitatively comparable
to the 3D results, i.e. these model potentials capture the essence
of the 3D process. This fact is in strong contrast with the poor results
of the conventional 1D soft-core and 1D Dirac-delta potentials, which
is caused mainly by their too weak and too strong binding force, respectively.

The graphs of the improved soft-core Coulomb potential are clearly
at the closest to the 3D reference in most of these cases, i.e. this
potential provides the quantitatively best model of the 3D case, despite
that its ground state density is not the exact reduced density of
the 3D case. The results of our numerical density-based model potential
are somewhat less close to the 3D reference. Although these simulations
start from the exact reduced density of the 3D case, the electron
is somewhat stronger bound to the ion-core than optimal. The results
obtained using the improved regularized Coulomb potential are very
close to those of the density-based model potential, but the former
potential is even somewhat stronger than needed.

In a typical strong-field simulation, the ground state population
loss $g(t)$ is close to the probability of ionization. Due to the
presence of the transverse degrees of freedom in 3D, it is then reasonable
that the $g(t)$ values are somewhat larger in a 3D simulation than
in 1D. Note that the $g(t)$ curves of the 1D simulations follow very
well the 3D reference curve in accordance with this.

The lack of the transverse degrees of freedom affects the $\left\langle v_{z}\right\rangle (t)$
curves of the 1D simulations in a different way: These exhibit the
high-frequency oscillations with larger amplitude than the 3D reference
curve. This can be explained by taking into account that rescattering
on the ion-core is a much stronger factor in 1D, and that the integration
over the transverse directions decreases the effect of the 3D density
oscillations on the reduced mean values. We will analyze this in more
detail in the next subsection.

In order to demonstrate the capabilities of these novel 1D model potentials,
we selected the time-dependent dipole moment $\left\langle z\right\rangle (t)$
to present the results of 4 different scenarios in Fig. \ref{fig:SinPulse3_F005_MeanZ}
and Fig. \ref{fig:SinPulse3_F015_NE_SAE}. Since the curves corresponding
to the density-based model potential are very close to those corresponding
to the improved regularized 1D Coulomb potential, we do not plot the
$\left\langle z\right\rangle (t)$ of this latter potential in all
of our Figures.

In Fig. \ref{fig:SinPulse3_F005_MeanZ} (a) we plot our simulation
results for hydrogen, now with a weaker field of $F=0.05$ which is
in the tunnel ionization regime of hydrogen, while Fig. \ref{fig:SinPulse3_F005_MeanZ}
(b) corresponds to a stronger field of $F=0.15$. Both of these figures
clearly show that the improved 1D soft-core Coulomb potential provides
the best results. Note that the change of $F$ in the above range
results in more than 2 orders of magnitude change in the peak value
of $\left\langle z\right\rangle (t)$.

Fig. \ref{fig:SinPulse3_F015_NE_SAE} (a) shows the results for a
Ne atom driven by a field of $F=0.15$. Here we model the 3D Neon
atom in the single active electron approximation \cite{lewenstein1994hhgtheory}
simply by setting the Coulomb-charge $Z_{{\rm Ne}}^{{\rm (SAE)}}=1.25929$
in order to match the ionization potential to the experimental value.
(For the improved regularized Coulomb potential $V_{0,{\rm M,C}}^{{\rm 1D}}$
we set $a_{{\rm Ne}}^{({\rm SAE)}}\approx0.26707525$ which yields
$E_{{\rm 0,M,C}}\approx-0.792905$.)

The accuracy of these 1D results is somewhat lower around the peak
and in the last half-period of the laser pulse than in the case of
hydrogen, and the improved soft-core Coulomb potential performs considerably
better in overall than the two other model potentials. By changing
the Coulomb charge $Z$ within a reasonable range in order to model
different noble gas atoms, we have obtained similarly accurate results.

Fig. \ref{fig:SinPulse3_F015_NE_SAE} (b) shows $\left\langle z\right\rangle (t)$
for a hydrogen atom, now driven by a longer laser pulse of shorter
carrier wavelength, corresponding to the parameters $T=80$, $F=0.1$,
and $N_{\text{Cycle}}=6$. The 1D model potentials work similarly
accurately for this longer laser pulse as in the case presented in
Fig. \ref{fig:SinPulse3_F010_MeanZ} (b), until the recollisions with
the ion-core gradually decrease the match between the 1D and 3D cases
in the last 2 periods of the pulse.

Our density-based 1D model potential and both of the improved 1D model
potentials exhibit an impressive improvement in the accuracy of the
low-frequency response of typical strong-field processes, in contrast
to the two conventional model potentials. These results are even more
convincing if we take into account that $\left\langle z\right\rangle (t)$,
$\sigma_{z}(t)$ and $g(t)$ are very sensitive to almost any change
in the physical parameter values.

\subsection{High order harmonic spectra}

In strong-field physics, the accurate computation of the high order
harmonic spectrum is especially important, because this represents
the highly nonlinear atomic response to the strong-field excitation,
with well-known characteristic features \cite{McPherson_JOSAB_1987_HHG,Ferray_JPhysB_1988_HHG,Harris_OptCom_1993_HHG_Atto,Krausz_RevModPhys_2009_Attosecond_physics,gombkoto2016quantoptichhg}.
Besides the high-order harmonic yield, the suitable phase relations
enable to generate attosecond pulses of XUV light \cite{Farkas_PhysLettA_1992_AttoPulse,Paul_Science_2001_Atto_pulsetrain,hentschel2001attosecond,drescher2002attosecond,kienberger2002attosecond,carrera2006attoxuv,sansone2006attosecondisolated}.

In Fig. \ref{fig:SinPulse3_Four_F010} (a), we plot the power spectrum
of the dipole acceleration (see Eq. \ref{eq:z_spectrumpow}) for the
parameters corresponding to Figs. \ref{fig:SinPulse3_F010_MeanZ}
and \ref{fig:SinPulse3_F010_Proj}.

In agreement with the previous subsection, the power spectra obtained
using the 1D model potentials agree very well with the 3D reference
simulation result up to the 5th harmonic. For higher frequencies,
the 1D spectra gradually deviate and give 1-2 orders of magnitude
larger values than the 3D reference values. The explanation given
for the oscillations of the $\left\langle v_{z}\right\rangle (t)$
curves in Fig. \ref{fig:SinPulse3_F010_Proj} (b) applies also here:
1D simulations exaggerate the effect of the ion-core, mainly via rescattering,
while the effect of the 3D density oscillations weakens in the reduced
mean values obtained from the 3D simulation.

However, the structure of the spectra in Fig. \ref{fig:SinPulse3_Four_F010}
(a) is remarkably similar and the match of the spectral phase, shown
in Fig. \ref{fig:SinPulse3_Four_F010} (b), is very good, especially
in the higher frequency range, which is of fundamental importance
for isolated attosecond pulses. These inspired us to create a scaling
function which transforms the spectra obtained with the 1D simulation
to fit the 3D reference spectrum as correctly as possible. Since the
improved soft-core Coulomb potential \eqref{eq:pot1_sc_corr} gives
the best low-frequency results, we focus only on this model potential
in the following.

Examination of the ratio of the magnitudes of the 1D power spectrum
to the 3D power spectrum in our simulations with different parameters
revealed that the scaling function 
\begin{equation}
s(f)=\min\left(1+0.03\left(100f-1\right)^{2},1+\left|100f-1\right|\right)\label{eq:pot1_four_sc_convert}
\end{equation}
transforms the magnitude of the power spectra obtained using the improved
1D soft-core Coulomb potential to properly fit the corresponding 3D
power spectra. In Fig. \ref{fig:SinPulse3_Four_F010_Ph} (a) we plot
the scaled 1D power spectrum $p(f)/s(f)$ which gives a very good
match between the 3D and 1D results in the case of the improved soft-core
Coulomb potential. (Here and in the following figures we plot the
scaled power spectrum of the density-based 1D model potential for
completeness only.) In Fig. \ref{fig:SinPulse3_Four_F010_Ph} (b)
and Fig. \ref{fig:SinPulse3_Four_F015_NE} (a) and (b) we present
this comparison for three other scenarios, corresponding to the parameters
of Fig. \ref{fig:SinPulse3_F005_MeanZ} (b) and Fig. \ref{fig:SinPulse3_F015_NE_SAE}
(a) and (b), respectively. These plots clearly show that the scaling
function \eqref{eq:pot1_four_sc_convert} works very well also in
these cases.

\section{Discussion and conclusions}

The results presented in the previous section demonstrate that it
is possible to quantitatively model the true 3D quantum dynamics with
the help of the density-based 1D model potential $\widetilde{V}_{0,{\rm M}}^{{\rm 1D}}(z_{j})$
and the accordingly improved soft-core Coulomb potential $V_{0,{\rm M,Sc}}^{{\rm 1D}}(z)$.
The best results are obtained with the improved soft-core Coulomb
potential \eqref{eq:pot1_sc_corr} which is also very easy to use
numerically. This means that we can perform quantum simulations of
a single active electron atom driven by a strong linearly polarized
laser pulse during a couple of minutes and obtain a fairly accurate
low-frequency response and a reliable HHG spectrum with the help of
the scaling function \eqref{eq:pot1_four_sc_convert}. The simple
form of this scaling is based on the good agreement between the structure
and phase of the 1D and the 3D HHG spectra.

In achieving these results, the physical requirement about the 1D
and 3D ground state densities was the important starting idea. This
led to the construction of the density-based 1D model potential, which
then inspired the improved parametrization of the 1D soft-core Coulomb
potential with effective charge $\frac{1}{2}Z$. Both of these have
the same asymptotic tail which ensures that their ground state energy
is identical to that of the 3D system. The discretization of the density-based
1D model potential gave important lessons also about the numerical
aspects of non-differentiable 1D Coulomb-like potentials and the 1D
delta potential.

Considering the obvious differences between the 1D and the 3D quantum
dynamics and their effects, discussed already in connection with Figs.
\ref{fig:SinPulse3_F010_Proj} (b) and \ref{fig:SinPulse3_Four_F010}
(a), it is not surprising that the high-frequency response of these
1D simulations is much stronger than that of the corresponding 3D
case. The fact that the scaling function \eqref{eq:pot1_four_sc_convert}
has different frequency-dependence in the lower frequency domain than
in the higher frequency domain, and that this seems to be independent
of the other physical parameters, may hint at a deeper connection
between the true 3D quantum dynamics and its best 1D model given by
the improved soft-core Coulomb potential \eqref{eq:pot1_sc_corr}.

We expect that this improved soft-core Coulomb potential can be successfully
used as a building block also in the 1D model of somewhat larger atomic
systems, like a He atom, driven by a strong linearly polarized laser
pulse. The method of construction of the reduced density-based 1D
model potential could be used as well to create proper 1D model potentials
for strong-field simulation of simple molecules, like $\mathrm{H}_{2}^{+}$
or $\mathrm{H}_{2}$. 
\begin{acknowledgments}
The authors thank F.\ Bogár, G.\ Paragi and S.\ Varró for stimulating
discussions. Szilárd Majorosi was supported by the UNKP-17-3 New National
Excellence Program of the Ministry of Human Capacities of Hungary.
The project has been supported by the European Union, co-financed
by the European Social Fund, EFOP-3.6.2-16-2017-00005. This work was
supported by the GINOP-2.3.2-15-2016-00036 project. Partial support
by the ELI-ALPS project is also acknowledged. The ELI-ALPS project
(GOP-1.1.1-12/B-2012-000, GINOP-2.3.6-15-2015-00001) is supported
by the European Union and co-financed by the European Regional Development
Fund. 
\end{acknowledgments}

\appendix

\section{Comparable physical quantities in 1D\label{sec:comparable-physical-quantities}}

For completeness, we list here the physical quantities that we use
for characterizing the strong-field process, both in 1D and 3D.

From the 3D wave function we can calculate the 1D reduced density
as

\begin{equation}
\varrho_{z}^{{\rm 3D}}(z,t)=2\pi\int_{0}^{\infty}\left|\Psi^{{\rm 3D}}(z,\rho,t)\right|^{2}\rho\text{d}\rho.\label{eq:z_reduced_density3}
\end{equation}
In 1D this is

\begin{equation}
\varrho_{z}^{{\rm 1D}}(z,t)=\left|\Psi^{{\rm 1D}}(z,t)\right|^{2}.\label{eq:z_density1}
\end{equation}
We calculate the mean value of $z$ as 
\begin{equation}
\left\langle z\right\rangle (t)=\int_{-\infty}^{\infty}z\varrho_{z}(z,t)\text{d}z,\label{eq:z_mean_z}
\end{equation}
the standard deviation of $z$ as 
\begin{equation}
\sigma_{z}(t)=\sqrt{\left\langle z^{2}\right\rangle (t)-\left\langle z\right\rangle ^{2}(t)},\label{eq:z_dev_z}
\end{equation}
the mean value of the $z$-velocity and the $z$-acceleration using
the Ehrenfest theorems as 
\begin{equation}
\left\langle v_{z}\right\rangle (t)=\frac{\partial\left\langle z\right\rangle }{\partial t},\,\,\,\,\,\,\,\left\langle a_{z}\right\rangle (t)=\frac{\partial\left\langle v_{z}\right\rangle }{\partial t},\label{eq:z_mean_vz}
\end{equation}
in both the 3D and the 1D cases. It is also interesting to determine
the ground state population loss 
\begin{equation}
g(t)=1-\left|\left\langle \Psi(0)|\Psi(t)\right\rangle \right|^{2},\label{eq:proj_ground_loss}
\end{equation}
even though this refers to the population losses of two completely
different states in 1D and 3D.

We calculate the spectrum from the dipole acceleration $\left\langle a_{z}\right\rangle $,
and then the power spectrum as 
\begin{equation}
p(f)=\left|\mathcal{F}\left[\left\langle a_{z}\right\rangle \right](f)\right|^{2}\label{eq:z_spectrumpow}
\end{equation}
where $\mathcal{F}$ denotes the Fourier transform and $f$ is its
frequency variable.

\section{Accuracy of the numerical inversion\label{sec:appendix_accuracy}}

\subsection{Density-based 1D model potential}

We stated previously that the numerical construction \eqref{eq:pot1_model_num}
yields the exact numerical eigensystem of that potential, but that
does not give us the whole picture about how numerically accurate
the construction really is. If we look at the eigenenergy of its respective
first excited state calculated with $\Delta z=0.2$ we see that it
is 4-5 digit accurate, but that alone does not determine the usefulness
in strong-field simulations. To get the whole picture, we performed
some numerical simulations using the atomic potential \eqref{eq:pot1_model_num}
and a 3-cycle laser pulse of form \eqref{eq:sim_sinpulse3_E} with
$F=0.1$ with different $\Delta z$ parameters. We subtracted from
them the results of a very accurate reference numerical solution using
the analytical potential \eqref{eq:pot1_model} with $\Delta z=0.0001$,
which gave us information about the (approximate) numerical errors
of the construction.

\begin{figure*}[t]
\begin{raggedright} \hspace{4.6cm}(a)\hspace{8.7cm}(b) 

\end{raggedright}

\includegraphics[width=1\columnwidth]{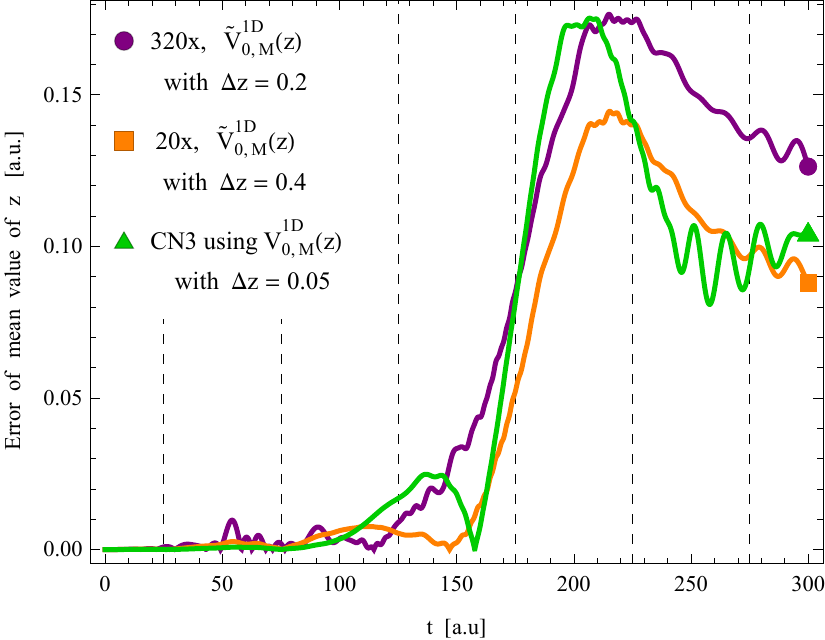}\hspace{0.5cm}\includegraphics[width=1\columnwidth]{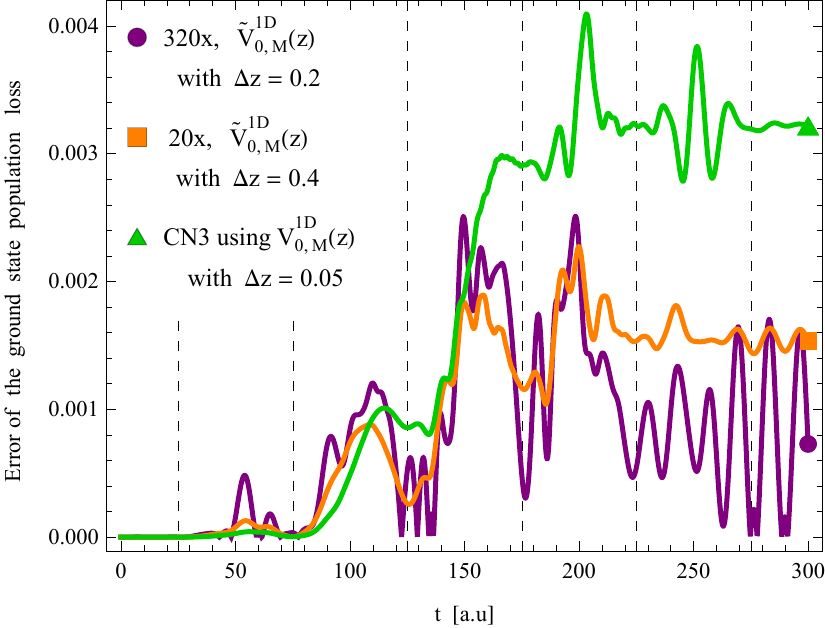}

\protect\protect\caption{Time-dependence of the numerical mean value errors $\left|\left\langle z\right\rangle (t)-\left\langle z\right\rangle _{\text{num,ref}}(t)\right|$
(a) and ground state population loss errors $\left|g(t)-g_{\text{num,ref}}(t)\right|$
(b) using different realizations of the density-based model potentials,
under the influence of the same external field with $F=0.1$, $N_{\text{Cycle}}=3$.
We plotted in purple and orange the results using the potential $\widetilde{V}_{0,{\rm M}}^{{\rm 1D}}$
from numerical inversion formula \eqref{eq:pot1_model_num} with $\Delta z=0.2$,
and $\Delta z=0.4$, respectively. Note that the values of these two
curves are magnified by a factor of 320 and 20 as indicated. For comparison,
we plotted in green the results directly using the analytic formula
\eqref{eq:pot1_model} as the atomic potential in an usual Crank-Nicolson
solution. }

\label{fig:SinPulse3_F010_Err_MeanZ} 
\end{figure*}

The results can be seen on Fig. \ref{fig:SinPulse3_F010_Err_MeanZ},
where we plotted the errors of mean values $\left\langle z\right\rangle (t)$
and the ground state population losses $g(t)$ compared to reference
versus time. We can see that if we decrease the spatial step $\Delta z$
of the inversion \eqref{eq:pot1_model_num} from 0.4 (orange) to 0.2
(purple) the error decreases approximately by a factor of 16, in the
case of both $\left\langle z\right\rangle (t)$ and $g(t)$. We verified
this using also other integrated quantities: we can clearly assert
that the numerical inversion \eqref{eq:pot1_model_num} is around
$\Delta z^{4}$ accurate i.e. it shows high order accuracy (required
that the kinetic energy operator is also at least $\Delta z^{4}$
accurate). To illustrate what this means, we also plotted the results
obtained by the usual 3-point finite difference Crank-Nicolson method
(CN3) using the analytical potential \eqref{eq:pot1_model} as the
atomic potential, which are known to be $\Delta z^{2}$ accurate.
We briefly note that we tested the direct use of \eqref{eq:pot1_model}
with our 11-point finite difference scheme but it was not any better,
also $\Delta z^{2}$ accurate (since the potential is not differentiable),
so we only plotted the results of the CN3 scheme in Fig. \ref{fig:SinPulse3_F010_Err_MeanZ}
with green lines. The accuracy of this method using $\Delta z=0.2$
is around 320 times better than the direct use of the analytical non-differentiable
potential with $\Delta z=0.05$. So in other words it requires $2^{6}$
more spatial gridpoints ($\Delta z\approx0.003$) than the numerical
inversion. Using the formula \eqref{eq:pot1_model_num} to numerically
represent the (nonsingular) model potentials is very efficient and
shows high order convergence.

\subsection{Delta potential}

\begin{figure*}[t]
\begin{raggedright} \hspace{4.6cm}(a)\hspace{8.7cm}(b) 

\end{raggedright}

\includegraphics[width=1\columnwidth]{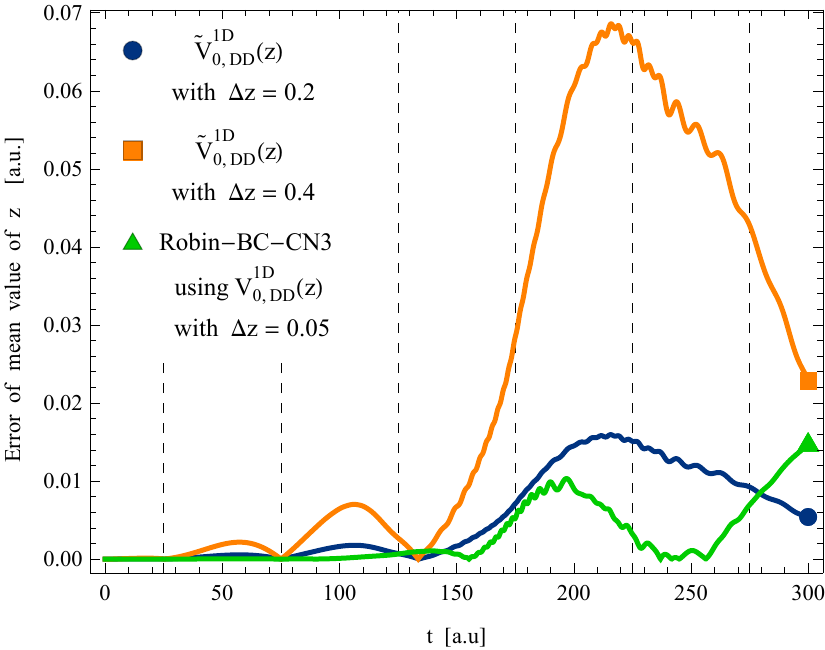}\hspace{0.5cm}\includegraphics[width=1\columnwidth]{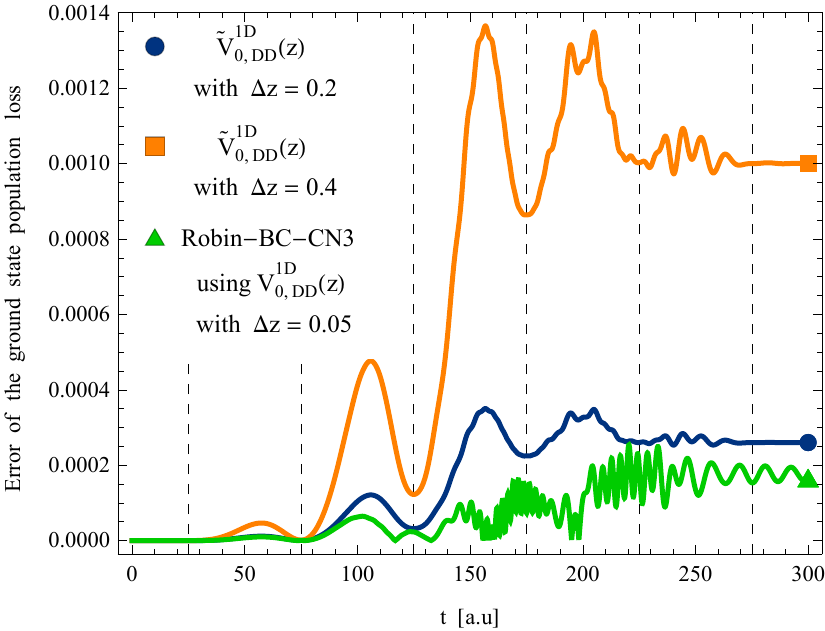}

\protect\protect\caption{Time-dependence of the numerical mean value errors $\left|\left\langle z\right\rangle (t)-\left\langle z\right\rangle _{\text{num,ref}}(t)\right|$
(a) and ground state population loss errors $\left|g(t)-g_{\text{num,ref}}(t)\right|$
(b) using different implementations of the Dirac-delta model potential,
under the influence of the same external field with $F=0.1$, $N_{\text{Cycle}}=3$.
We plotted in dark blue and orange the results using the potential
$\widetilde{V}_{0,{\rm DD}}^{{\rm 1D}}$ from numerical inversion
formula \eqref{eq:pot1_dd_num} with $\Delta z=0.2$, and $\Delta z=0.4$,
respectively. For comparison, we plotted in green the results using
the implementation in \cite{czirjak2013rescatterentanglement} that
uses the proper Robin boundary condition to represent the singularity
of \eqref{eq:pot1_dd}. }

\label{fig:SinPulse3_F010_Err_DD_MeanZ} 
\end{figure*}

In the following, we discuss the accuracy tests of the numerically
constructed potential \eqref{eq:pot1_dd_num} using strong-field simulations
with the same 3-cycle laser pulse of form \eqref{eq:sim_sinpulse3_E}
with $F=0.1$ and different $\Delta z$ parameters. For comparison
we use a properly implemented method from \cite{czirjak2013rescatterentanglement}
that uses the proper Robin boundary condition at $z=0$, which overrides
the Crank-Nicolson equations at that grid point. Its results are at
least $\Delta z^{2}$ accurate. We calculate the errors of the mean
values $\left\langle z\right\rangle (t)$ and the ground state population
losses $g(t)$ compared to a very accurate reference solution obtained
by this correct method (uses $\Delta z=0.001$). We can see the results
on Fig. \ref{fig:SinPulse3_F010_Err_DD_MeanZ}. Surprisingly, we can
observe that the errors of \eqref{eq:pot1_dd_num} with $\Delta z=0.2$
are actually not far from the errors of results obtained by the $\Delta z^{2}$
accurate proper method at $\Delta z=0.05$. If we decrease the $\Delta z$
step from 0.4 (orange) to 0.2 (dark blue) we see a factor 4 error
decrease: we can conclude that the non-singular construction \eqref{eq:pot1_dd_num}
is actually correct numerical representation, and converges with $\Delta z^{2}$
even for the singular delta potential. It is also of importance because
of the following: we can run simulations with singular potentials
using non-singular Hamiltonians, and the point of singularity is not
have to be on the spatial grid, it can even move. It has even more
interesting consequences in 2D or more, since there is no reason not
to work with the true singular Coulomb potentials.

In conclusion, it is a valid choice to define potentials using numerical
inversion from its ground state. It can provide a consistent and accurate
method with high order finite differences to represent our \eqref{eq:pot1_model}
non-singular and non-differentiable atomic potential in 1D, and even
achieve $\Delta z^{4}$ convergence. The method is robust enough to
provide $\Delta z^{2}$ convergence for the case of the singular 1D
delta potential using \eqref{eq:pot1_dd_num}.

\bibliographystyle{unsrtnat}
\bibliography{0Bibliography,2Bibliography,3Bibliography}

\begin{figure*}
\begin{raggedright} \hspace{4.6cm}(a)\hspace{8.7cm}(b) 

\end{raggedright}

\includegraphics[width=1\columnwidth]{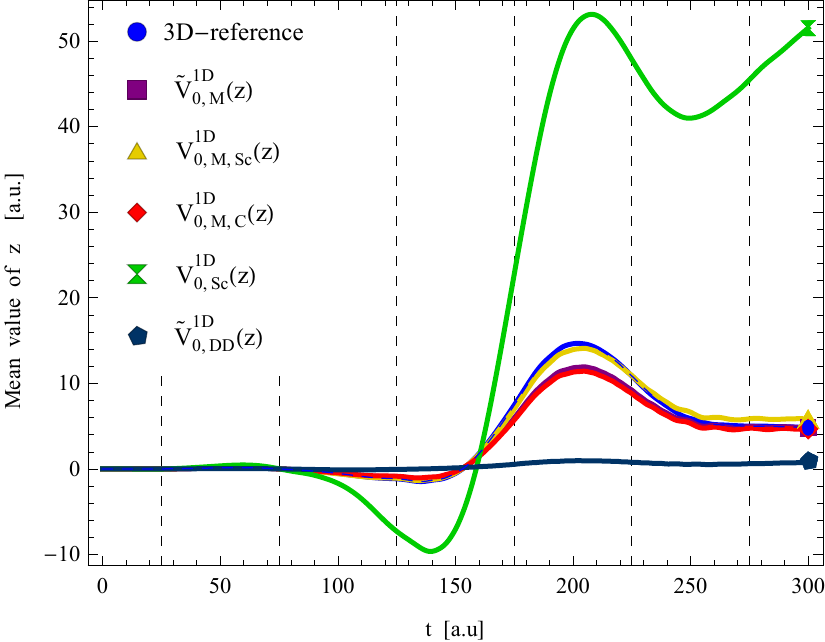}\hspace{0.5cm}\includegraphics[width=1\columnwidth]{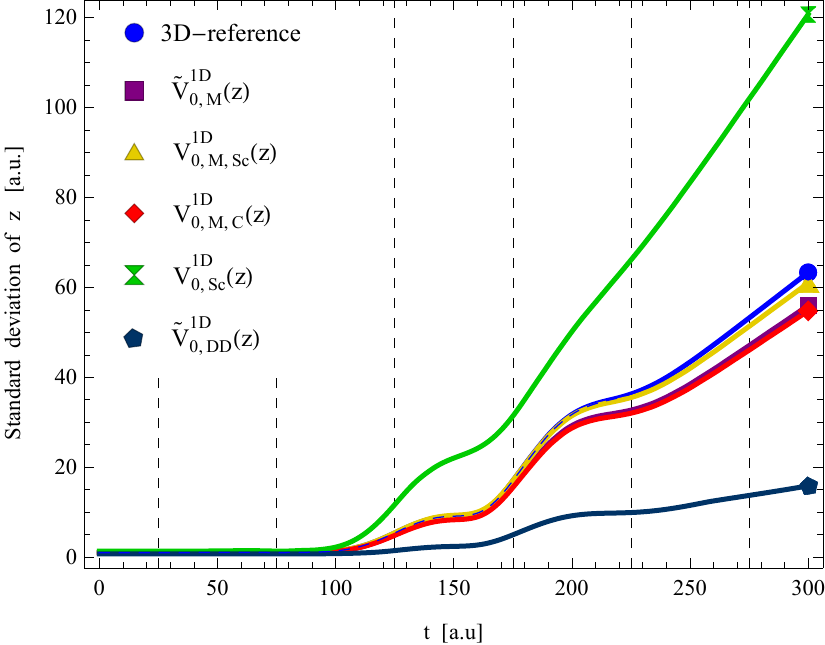}

\protect\protect\caption{Time-dependence of the mean values $\left\langle z\right\rangle (t)$
(a) and the standard deviations $\sigma_{z}(t)$ (b) using different
1D model potentials, under the influence of the same external field
with $F=0.1$, $N_{\text{Cycle}}=3$ and $T=100$. Results of the
corresponding 3D simulation are plotted in blue.}

\label{fig:SinPulse3_F010_MeanZ} 
\end{figure*}

\begin{figure*}
\begin{raggedright} \hspace{4.6cm}(a)\hspace{8.7cm}(b) 

\end{raggedright}

\includegraphics[width=1\columnwidth]{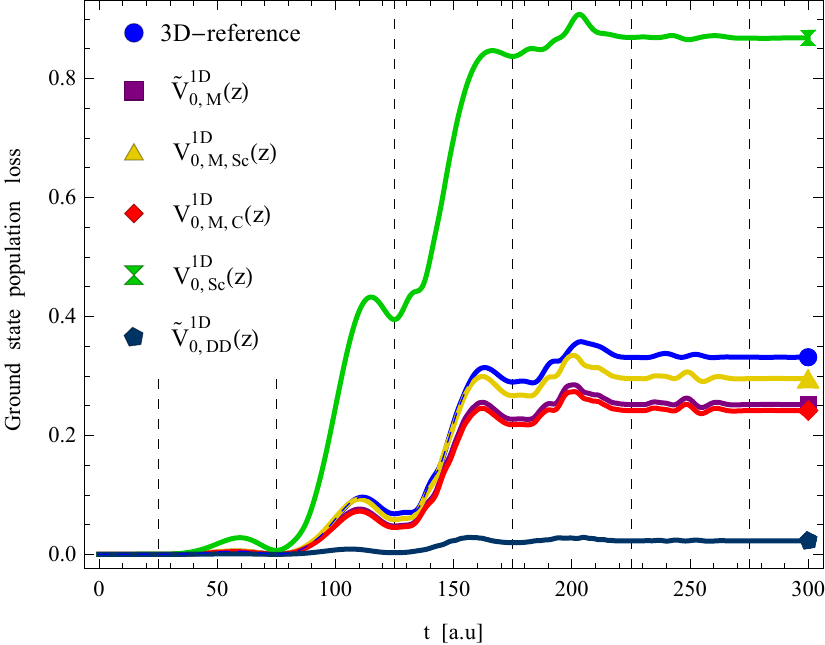}\hspace{0.5cm}\includegraphics[width=1\columnwidth]{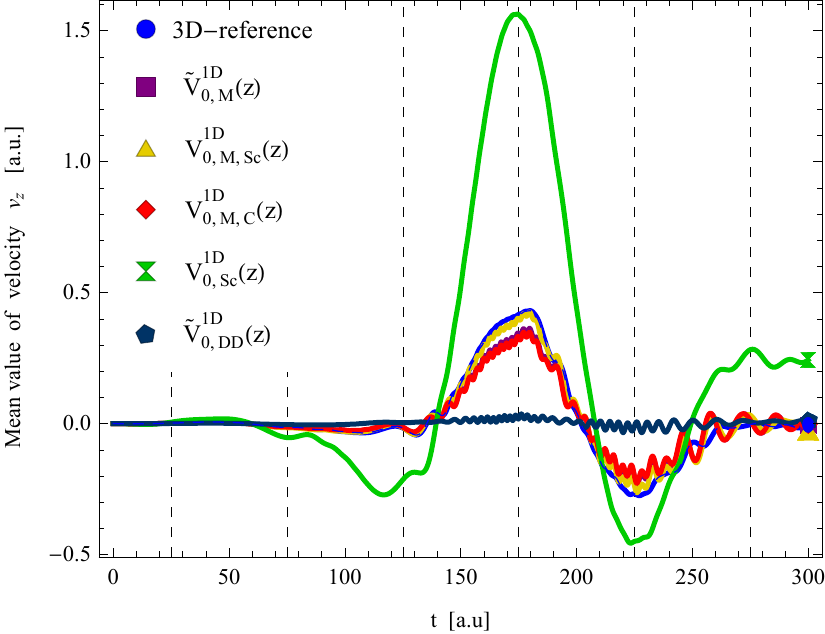}

\protect\protect\caption{Time-dependence of the ground state population loss $g(t)$ (a) and
the mean velocities $\left\langle v_{z}\right\rangle (t)$ (b) using
different 1D model potentials, under the influence of the same external
field with $F=0.1$, $N_{\text{Cycle}}=3$ and $T=100$. Results of
the corresponding 3D simulation are plotted in blue.}

\label{fig:SinPulse3_F010_Proj} 
\end{figure*}

\begin{figure*}
\begin{raggedright} \hspace{4.6cm}(a)\hspace{8.7cm}(b) 

\end{raggedright}

\includegraphics[width=1\columnwidth]{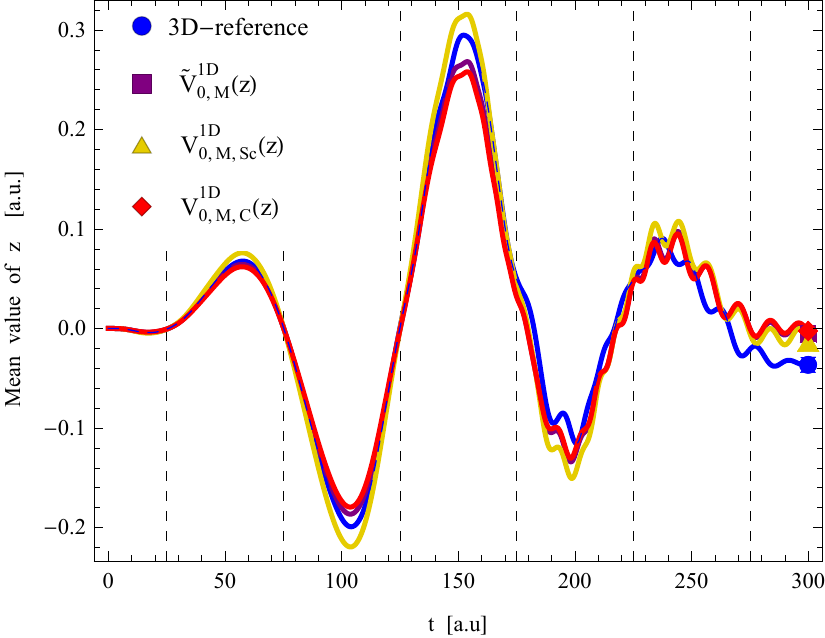}\hspace{0.5cm}\includegraphics[width=1\columnwidth]{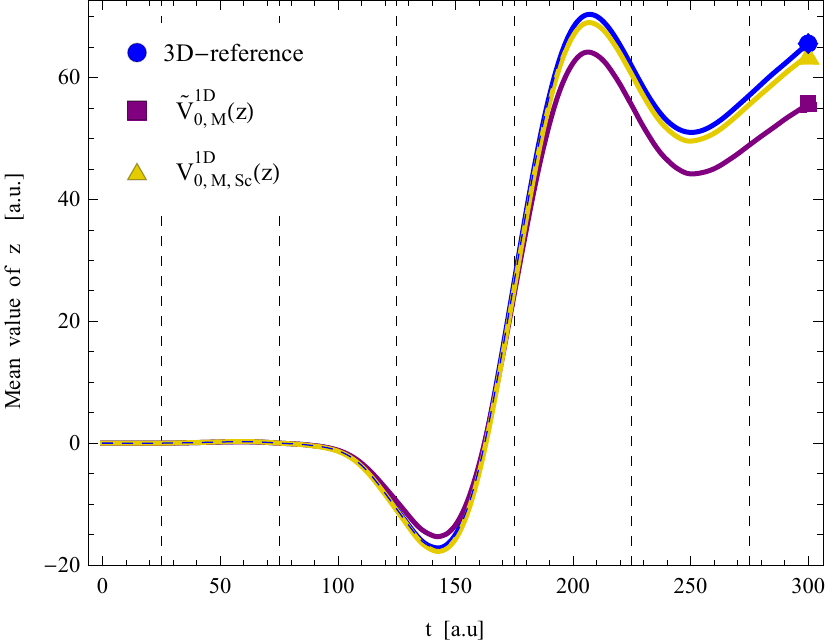}

\protect\protect\caption{Time-dependence of mean values $\left\langle z\right\rangle (t)$
using different 1D model potentials, under the influence of the external
field with $F=0.05$ (a) and $F=0.15$ (b), $N_{\text{Cycle}}=3$
and $T=100$. Results of the corresponding 3D simulations are plotted
in blue.}

\label{fig:SinPulse3_F005_MeanZ} 
\end{figure*}

\begin{figure*}
\begin{raggedright} \hspace{4.6cm}(a)\hspace{8.7cm}(b) 

\end{raggedright}

\includegraphics[width=1\columnwidth]{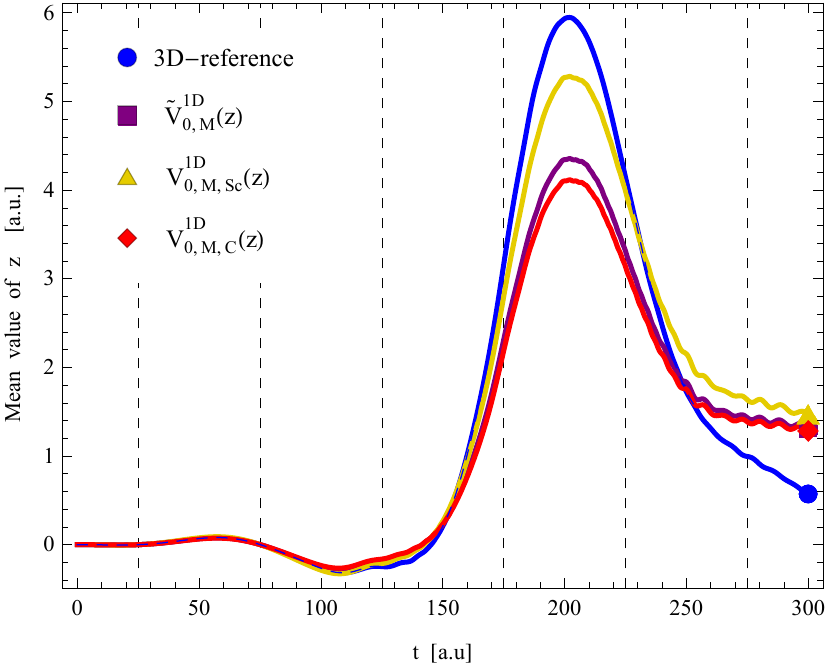}\hspace{0.5cm}\includegraphics[width=1\columnwidth]{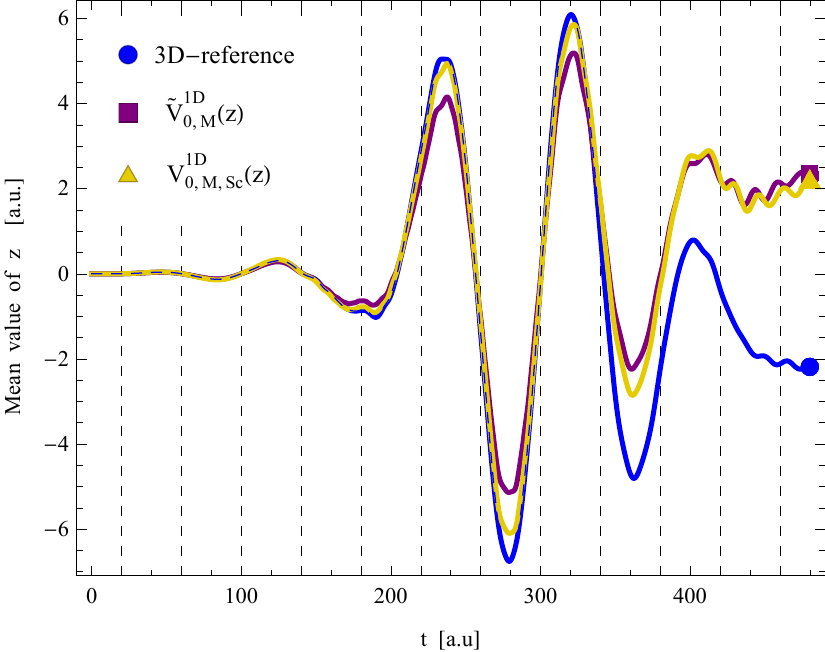}

\protect\protect\caption{Time-dependence of mean values $\left\langle z\right\rangle (t)$
using different 1D model potentials. Panel (a): single active electron
model of a neon atom with $Z_{{\rm Ne}}^{{\rm (SAE)}}=1.25929$, driven
by the external field with $T=100$, $F=0.15$, $N_{\text{Cycle}}=3$.
Panel (b): hydrogen with $Z=1$ using the driven by the external field
with $T=80$, $F=0.1$, $N_{\text{Cycle}}=6$. Results of the corresponding
3D simulations are plotted in blue.}

\label{fig:SinPulse3_F015_NE_SAE} 
\end{figure*}

\begin{figure*}
\begin{raggedright} \hspace{4.6cm}(a)\hspace{8.7cm}(b) 

\end{raggedright}

\includegraphics[width=1\columnwidth]{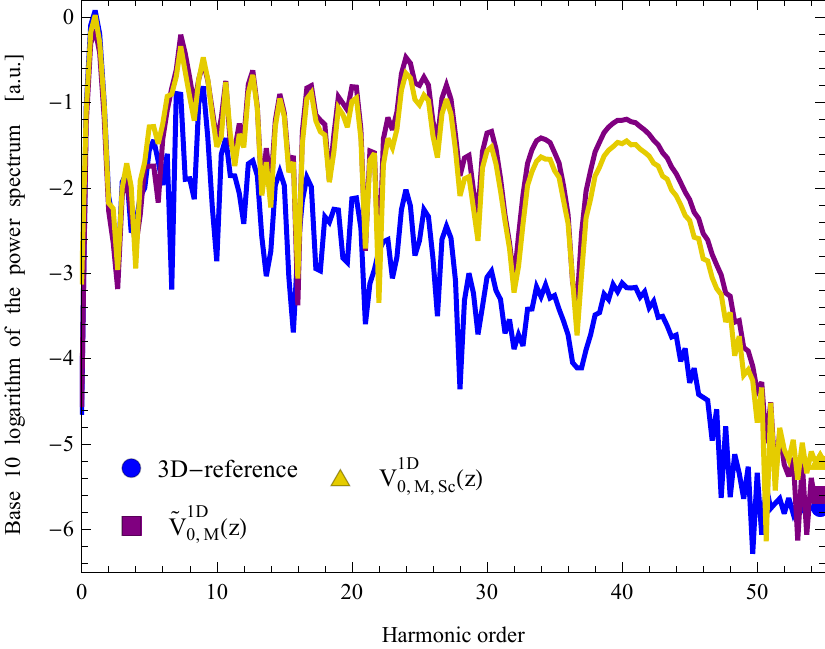}\hspace{0.5cm}\includegraphics[width=1\columnwidth]{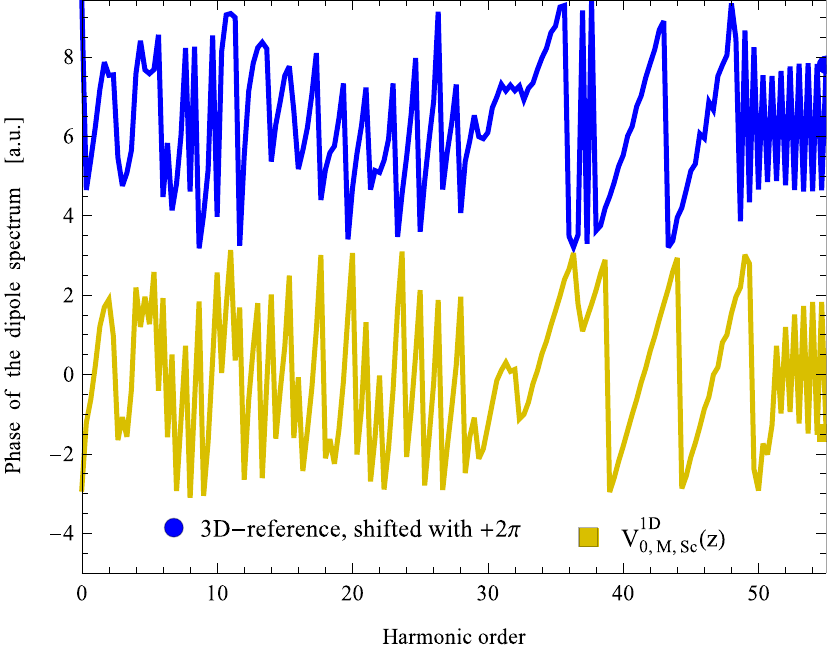}

\protect\protect\caption{Panel (a): Logarithmic plot of the power spectra vs. the harmonic
order, i.e. $p(nf_{1})$ (where $f_{1}=1/T=0.01$ a.u. is the fundamental
frequency). Panel (b): Phase of the dipole acceleration spectra vs.
the harmonic order (up shifted by $2\pi$ for the 3D case). We plot
the results for the density-based 1D model potential (purple) and
for the improved soft-core Coulomb potential (gold) in comparison
with the 3D reference (blue). The parameters $F=0.1$, $N_{\text{Cycle}}=3$,
$T=100$ and $Z=1$ are the same as for Figs. \ref{fig:SinPulse3_F010_MeanZ}
and \ref{fig:SinPulse3_F010_Proj}. }

\label{fig:SinPulse3_Four_F010} 
\end{figure*}

\begin{figure*}
\begin{raggedright} \hspace{4.6cm}(a)\hspace{8.7cm}(b) 

\end{raggedright}

\includegraphics[width=1\columnwidth]{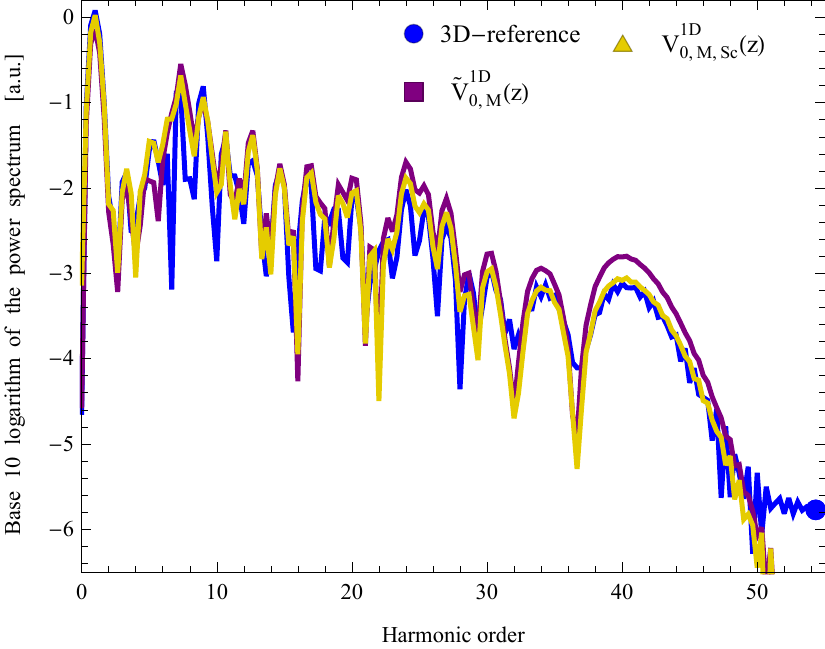}\hspace{0.5cm}\includegraphics[width=1\columnwidth]{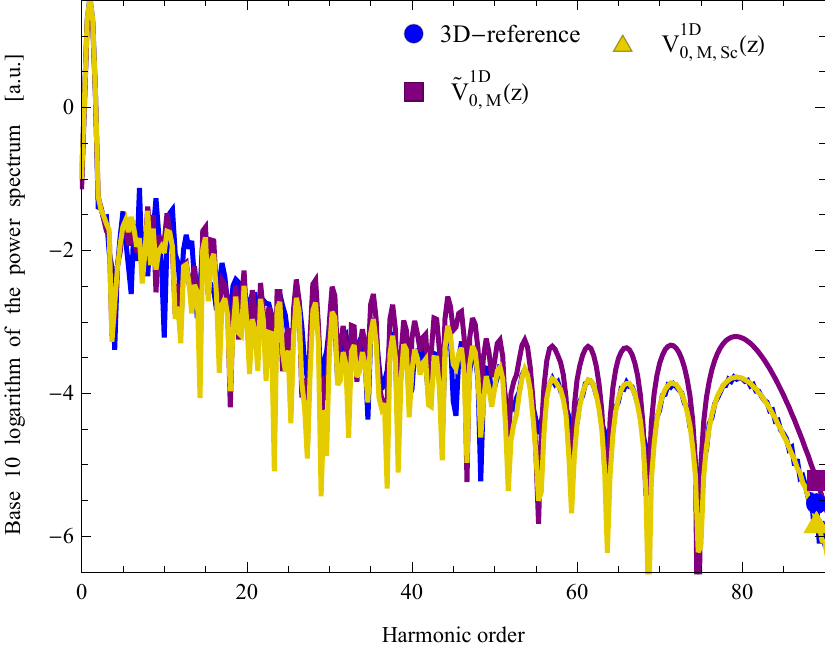}

\protect\protect\caption{Logarithmic plots of the scaled power spectra $p(nf_{0})/s(nf_{0})$
using the model systems of Fig. \ref{fig:SinPulse3_Four_F010} with
$F=0.10$ (a), $F=0.15$ (b), in comparison with the 3D reference
(blue). }

\label{fig:SinPulse3_Four_F010_Ph} 
\end{figure*}

\begin{figure*}
\begin{raggedright} \hspace{4.6cm}(a)\hspace{8.7cm}(b) 

\end{raggedright}

\includegraphics[width=1\columnwidth]{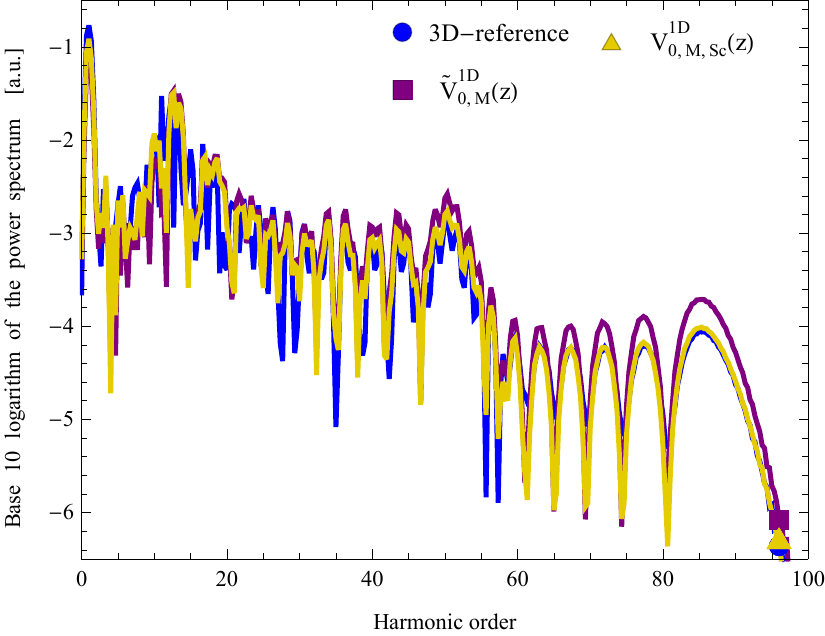}\hspace{0.5cm}\includegraphics[width=1\columnwidth]{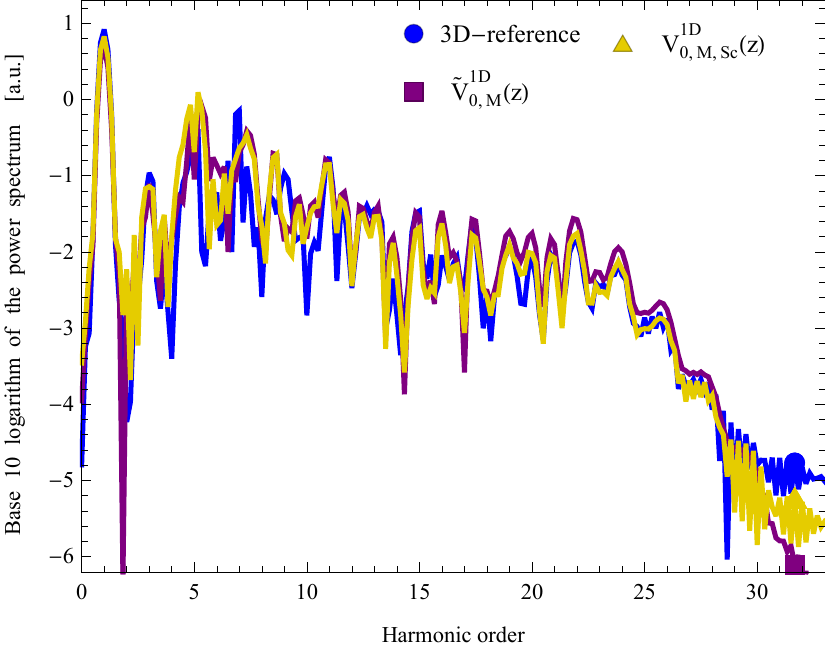}

\protect\protect\caption{Logarithmic plots of the scaled power spectra $p(nf_{0})/s(nf_{0})$
obtained using the density-based 1D model potential (purple) and the
improved soft-core Coulomb potential (gold), in comparison with the
3D reference (blue). Panel (a): single active electron model of a
neon atom with $Z_{{\rm Ne}}^{{\rm (SAE)}}=1.25929$ driven by the
external field with $T=100$, $F=0.15$, $N_{\text{Cycle}}=3$. Panel
(b): hydrogen with $Z=1$, driven by the external field with $T=80$,
$F=0.1$, $N_{\text{Cycle}}=6$.}

\label{fig:SinPulse3_Four_F015_NE} 
\end{figure*}

\end{document}